\pdfoutput=1
\documentclass[letterpaper,12pt]{article}

\usepackage{times}
\usepackage{helvet}
\usepackage{courier}
\usepackage[utf8]{inputenc}
\usepackage[english]{babel}
\usepackage{amsmath}
\usepackage{amsthm}
\usepackage{amssymb}
\usepackage{url}
\usepackage{graphicx}
\usepackage{array}
\usepackage{float}

\usepackage{cite}
\usepackage{amsmath,amssymb,amsfonts}
\usepackage{algorithmic}
\usepackage{graphicx}
\usepackage{subcaption}
\usepackage{textcomp}
\usepackage{xcolor}
\def\BibTeX{{\rm B\kern-.05em{\sc i\kern-.025em b}\kern-.08em
    T\kern-.1667em\lower.7ex\hbox{E}\kern-.125emX}}

\usepackage{todonotes} 
\usepackage{tikz}
\usepackage{multirow}
\usepackage{makecell} 
\usepackage{enumerate}
\usepackage{url}
\usepackage{authblk}

\newtheorem{definition}{Definition}
\usetikzlibrary{positioning}
\usetikzlibrary{chains}
\usetikzlibrary{shapes}
\usetikzlibrary{calc}
\usetikzlibrary{arrows.meta}
\usetikzlibrary{overlay-beamer-styles}
\usetikzlibrary{decorations.pathreplacing}

\newcommand{\Graph}{\mathcal{G}}
\newcommand{\EdgeSet}{E}
\newcommand{\NodeSet}{V}
\newcommand{\HIN}{\mathcal{H}}
\newcommand{\NodeLabels}{\mathcal{A}}
\newcommand{\EdgeLabels}{\mathcal{R}}
\newcommand{\NodeMap}{\varphi}
\newcommand{\EdgeMap}{\psi}
\newcommand{\Users}{\mathcal{U}}
\newcommand{\Items}{\mathcal{I}}

\newcommand{\diver}{\mathcal{P}}
\newcommand{\coldiv}[1]{\mathcal{P}_{\text{Col}}(#1)}
\newcommand{\midiv}[1]{\mathcal{P}_{\text{MI}}(#1)}
\usepackage{xspace}
\newcommand{\eg}{{\em e.g.,}\xspace}
\newcommand{\ie}{{\em i.e.,}\xspace}

\begin{document}

\title{Testing the Impact of Semantics and Structure on Recommendation Accuracy and Diversity}

\author[1]{Pedro Ramaciotti Morales}
\author[2]{Lionel Tabourier}
\author[3]{Rapha\"el Fournier-S'niehotta}

\affil[1]{M\'edialab, Sciences Po, Paris, France}
\affil[2]{Sorbonne Universit\'e, CNRS, LIP6, F-75005 Paris, France}
\affil[3]{CEDRIC, CNAM, Paris, France}

\date{}

\maketitle

\begin{abstract}
The Heterogeneous Information Network (HIN) formalism is very flexible and enables complex recommendations models.
We evaluate the effect of different parts of a HIN on the accuracy and the diversity of recommendations, then investigate if these effects are only due to the semantic content encoded in the network.
We use recently-proposed diversity measures which are based on the network structure and better suited to the HIN formalism.
Finally, we randomly shuffle the edges of some parts of the HIN, to empty the network from its semantic content, while leaving its structure relatively unaffected.
We show that the semantic content encoded in the network data has a limited importance for the performance of a recommender system and that structure is crucial.
\end{abstract}


\section{Introduction}
\label{sec:intro}

A Recommender System (RS) helps users in selecting items from large sets when their size makes individual consideration impractical or even impossible~\cite{ricci2011introduction}.
To propose relevant recommendations to users, a RS may process previous choices made by users, item/user meta-data, and contextual information.
Initially, the performances of RS were measured with the so-called utility or accuracy metrics~\cite{herlocker2004evaluating}, which evaluate, for example, the error committed when predicting the rating a user would give to a specific item. 
Later, other properties of recommendations beyond accuracy were suggested, because users also wish for surprising recommendations and thus should be proposed diverse items~\cite{mcnee2006being}.
Moreover, diversifying recommendations is also desirable to guarantee that people are not shut in a small subset of homogeneous recommendations, a phenomenon known as a filter bubble~\cite{pariser2011filter}.

RS models are often separated between content-based filtering and collaborative filtering approaches.
Many recent RS tend to use hybrid methods that mix both approaches~\cite{burke2002hybrid}.
Among these, methods relying on the Heterogeneous Information Network (HIN) framework are gaining popularity~\cite{sun2011pathsim}.
HINs are graph-based representations with different types of node: some represent users, others represent items, and a new type of node (or layer) is added for each kind of information in the system~\cite{burke2014hybrid}.
Then nodes of different types are connected with links representing a specific type of relation, such as \textit{a user selected an item}, or \textit{an item belongs to a given genre}, etc.
A simple example consists in {\em readers} linked to {\em books} they read, which are in turn linked to {\em genres} (categories).

In this work, we want to evaluate how different parts of a HIN impact the diversity and the accuracy of the recommendations a RS based on this HIN may produce.
Using a RS derived from Pathsim~\cite{sun2011pathsim}, which relies on random walks on HINs, we may gradually include different parts of the HIN into the RS in a controlled way.
These different parts encode different meta-data on users and items: some nodes may represent the demographic group of users (say, their gender), others may represent a category of items (e.g., drama or comedy for movies), etc.  
Most current diversity measures may not benefit from the full structural information of networks, since they are defined only with the output of the recommendation process (the lists of recommended items and their features)~\cite{kunaver2017diversity,silveira2019good}. Thus, our evaluation of diversity relies on a framework especially suited to HINs~\cite{ramaciotti2020measuring}.

Another important goal of our work is to measure whether the performance of a HIN-based RS is mostly due to the nature of the semantic information encoded, or to a {\em structural shortcut effect}.
A HIN representation may contain connections between items or users who may not share a lot of common features.
For instance, two users might have different tastes while living in the same city and a HIN representation would connects them through their zip code. 
A HIN-based RS would use these ``shortcuts'' to recommend items otherwise unreachable, affecting diversity.
We address this problem of balance between structural and semantic information by using a \textit{configuration model}~\cite{newman2018networks} of our HIN. Randomly shuffling edges in a given part of the HIN (while keeping the degree distribution unaffected), we separate the effects of those semantic and structural information.

To summarize, the research questions addressed are: 
\begin{itemize}
  \item[Q1:] How to test the effect of different parts of a HIN on the accuracy and the diversity of recommendations?, and 
  \item[Q2:] Are the effects of different parts of a HIN on accuracy and diversity due to the semantic content encoded in them?
\end{itemize}

This article is structured as follows.
In Section~\ref{sec:related_work}, we discuss the related work.
In Section~\ref{sec:preliminaries}, we provide the definitions of concepts used in this paper, in particular the formal definition of a HIN and suited diversity measures.
We also expose the design of our HIN-based RS model.
In Section~\ref{sec:experiments}, we detail a series of experiments which correspond to one of our contributions:  
\begin{enumerate}
  \item we provide the first application of {\em network diversity measures} to recommendations computed on HINs (Sec.~\ref{subsec:exploring_diversities}). 
  \item we use a simple, well-known procedure  to compute recommendations on HINs~\cite{sun2011pathsim}, and test it on the datasets to answer question Q1. We show that depending on the information included, such a method can yield accurate and diverse recommendations (Sec.~\ref{subsec:structure_accuracy_diversity}). 
  \item finally, we use edge shuffling of some parts of a HIN to show that it is not always possible to explain the performance of recommendations by the semantic information encoded in HINs, thus answering question Q2 (Sec.~\ref{subsec:acc_diversity_config_model}). To the best of our knowledge, this is the first work to investigate how the global structure of a HIN affects accuracy and diversity of recommendations.
\end{enumerate}


\section{Related Work}
\label{sec:related_work}

HIN-based RSs are rapidly developing into a subdomain of information filtering and recommendation~\cite{burke2014hybrid,shi2016survey}.
HINs enable mixing past choices of users with semantic information about items and users in a unified graph-based representation.
In a HIN, if a set of users are linked to books they have chosen, and these books linked to genres, for example, a HIN-based RS may exploit the collective choices made by users (the set of links representing choice), and the genres of books chosen by users.
Recommendations based on content (recommending books of previously chosen genre), but also on collective choice (recommending books chosen by users that have similar taste) are then possible.
The HIN formalism has been shown to perform well and to improve interpretability~\cite{shi2015semantic,shi2018heterogeneous}.
Moreover, HINs can deal with cross-domain recommendation, often useful in cold-start situations~\cite{tang2012cross}, and can exploit social relations between users~\cite{yang2012circle}.
They also allow predictions based on implicit feedback~\cite{yu2014personalized}.

RS evaluation usually focuses on \textit{accuracy}: measuring the error on predicted ratings with metrics such as RMSE or MAE, or measuring the error committed on items selected by users using metrics such as precision, recall, or F1-score.
However, almost two decades ago, it was acknowledged that properties beyond accuracy play an important role in user satisfaction~\cite{herlocker2004evaluating,mcnee2006being}.
These properties aim at grasping how novel or surprising a recommendation is~\cite{herlocker2004evaluating} or how diverse the proposed recommendation list is~\cite{ziegler2005improving}.
There is little consensus on the naming of these desirable properties, but they may be collectively referred to as \textit{diversity} measures, indicating a degree of (dis-)similarity between recommended items.
Some authors consider two items similar if they are chosen by the same users~\cite{hurley2011novelty,zhang2002novelty}, while others consider items similar if they are of the same type~\cite{ziegler2005improving}.
These notions are often called \textit{novelty} or \textit{intra-list similarity} .
Some works focus on the improbability, or the \textit{surprisal}, of a recommendation considering the users' past choices, while others focus on \textit{personalization}, {\em i.e.,} the degree to which recommendations are different from one user to another~\cite{zhou2010solving}.
Measuring diversity is also important for specific applications, such as detecting context changes~\cite{lhuillier2016new} or measuring phenomena related to diversity loss, such as filter bubbles and echo chambers~\cite{pariser2011filter}.
An extensive survey recently compiled diversity measures for RS~\cite{kunaver2017diversity}.

In information filtering, searches in graphs sometimes call to some notion of diversity on node rankings (i.e., item rankings)~\cite{tong2011diversified,li2012scalable}.
While many studies consider diversity with respect to some node labeling, few exploit the variety of paths formed by relations between nodes of different types~\cite{kunaver2017diversity}.
Dubey et al.~\cite{dubey2011diversity} use, as we do in this article, the notion of random walk along these paths, which are usually called \textit{meta-paths} in HINs.
Furthermore, some works use the notion of random walk in relation to diversity~\cite{nandanwar2018fusing,jiang2019degenerate}, although not as a measure to evaluate resulting recommendations of RSs in HINs.
While there are theoretical propositions of diversity measures adapted to HIN~\cite{ramaciotti2020measuring}, there are no available experiments that illustrate their applicability.


\section{Heterogeneous Information Networks and Diversity}
\label{sec:preliminaries}

In this section we give definitions and notations about HINs that are used in the rest of this paper.
By combining these definitions with random walks on graphs, we are able to define the diversity measures that are then exploited in Section~\ref{sec:experiments}.

\subsection{Heterogeneous Information Networks}
\label{sec:hins}

\begin{definition}[Heterogeneous Information Network \cite{shi2017heterogeneous}]
A Heterogeneous Information Network (HIN) $\HIN=(\Graph,\NodeLabels,\EdgeLabels,\NodeMap,\EdgeMap)$ is a directed graph $\Graph=(\NodeSet,\EdgeSet)$
with mapping functions $\NodeMap:\NodeSet\rightarrow \NodeLabels$ and $\EdgeMap:\EdgeSet\rightarrow \EdgeLabels$, where $\NodeLabels$ and $\EdgeLabels$ are respectively node and edge label sets.
\end{definition}

Slightly simplifying notations, we refer to a group of nodes or edges by its label.
Edge labels in $\EdgeLabels$ represent relations between entities of different object groups in $\NodeLabels$.

\begin{definition}[Heterogeneous Information Network Schema \cite{shi2017heterogeneous}]
The network schema of a HIN $\HIN=(\Graph,\NodeLabels,\EdgeLabels,\NodeMap,\EdgeMap)$ is a directed graph, the nodes of which are the types $\NodeLabels$, and the edges are the relations of $\EdgeLabels$.
\end{definition}

\begin{figure}[!h]
  \centering
\begin{tikzpicture}[{
  > = {Stealth [inset = 0pt, length = 10pt, angle' = 30, round]},
  vertexset/.style = {minimum width = 0.5cm, minimum height = 0.5cm, inner sep = 0pt},
  vertex/.style = {minimum width = 0.75cm, minimum height = 0.5cm, inner sep = 0pt, draw, circle},
  marked/.style = {draw = blue, color = blue, line width = (4*#1), text = blue},
  edge/.style={->}
}]
\small
  \node [vertex] (u) {U};
  \node [vertex, right = 1.5cm of u] (i) {I};
  \node [vertex, right = 1.5cm of i] (Ty) {Ty};

  \draw[->] (u) to[out=0, in=180, edge node={node [midway,above] {$R_{\text{rates}}$}}] (i) ;
  \draw[->] (i) to[out=0, in=180, edge node={node [midway,above] {$R_{\text{Ty}}$}}] (Ty);
\end{tikzpicture}
\caption{\label{fig:smallest} HIN schema of a simple case where users (U) rate items (I) and items belong to types (Ty), thus defining edge labels $R_{rates}$ and $R_{Ty}$.
}
\end{figure}
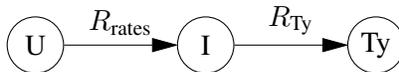

A {\em HIN schema} (or simply {\em schema}) is a graphical and summarized representation of a HIN, which shows the groups of nodes and links.
Figure~\ref{fig:smallest} shows a basic example of HIN schema, corresponding to a
platform where users would rate items and items belong to predefined categories.
Other examples can be found in Section~\ref{subsec:datasets}.

\begin{definition}[Link Group]
\label{def:link_group}
A link group, represented by an edge in the {\em HIN schema} $R \in \EdgeLabels$, is the subset
 of edges in $\EdgeSet$ that are of type $R$,
linking source object group $S\in \NodeLabels$ and target object group $T\in \NodeLabels$.
It is denoted simply by $R$, and its inverse (made of the reversed edges of a link) by $R^{-1}$.
\end{definition}

\begin{definition}[Meta-path \cite{shi2017heterogeneous}]
A meta-path of a HIN is a path on its schema, and it is denoted by $\Pi=A_1 \xrightarrow[]{R_1} A_2 \xrightarrow[]{R_2} ... \xrightarrow[]{R_l} A_{l+1}$,
or more simply by $\Pi = R_1 R_2 \cdots R_{l}$.
\end{definition}

In the example of Figure~\ref{fig:smallest}, $\Pi = R_{rates} R_{Ty} $ is a meta-path connecting users to the types of items that they have rated.

\subsection{Random Walks and Meta-Path Diversities on HINs}
\label{sec:random_walks}

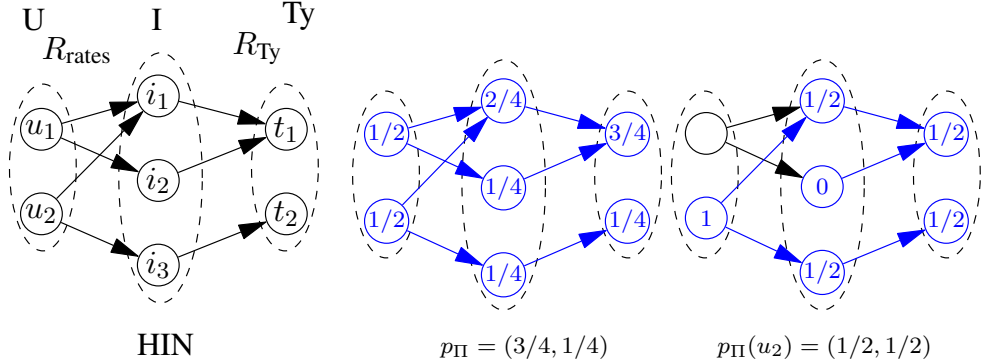
\begin{figure}[!h]
  \resizebox{.98\columnwidth}{!}{
\begin{tikzpicture}[{
  > = {Stealth [inset = 0pt, length = 10pt, angle' = 30, round]},
  vertexset/.style = {minimum width = 0.5cm, minimum height = 0.5cm, inner sep = 0pt},
  vertex/.style = {minimum width = 0.25cm, minimum height = 0.5cm, inner sep = 0pt, draw, circle},
  marked/.style = {draw = blue, color = blue, line width = (4*#1), text = blue},
  edge/.style={->}
}]
\small
  \node (center) {};

  \node [below right = 2.9cm and -2.5cm of center,text width=4cm, align=center] (hin) {HIN};

  \node [above left = 0.2cm and 1.4cm of center] (Uname) {U};
  \node [above left = -0.2cm and 0.6cm of center] (Rrate) {$R_\text{rates}$};
  \node [vertex, below left = 0.4cm and 1.4cm of center] (u1) {$u_1$};
  \node [vertex, below = 0.5cm of u1] (u2) {$u_2$};
  \draw [dashed] (-1.7,-1.175) ellipse (0.4cm and 1cm);
  \node [above left = 0.2cm and 0cm of center] (Iname) {I};
  \node [vertex, below left = 0.0cm and 0cm of center] (i1) {$i_1$};
  \node [vertex, below = 0.5cm of i1] (i2) {$i_2$};
  \node [vertex, below = 0.5cm of i2] (i3) {$i_3$};
  \draw [dashed] (-0.29,-1.3) ellipse (0.5cm and 1.5cm);
  \node [above right = 0.2cm and 0.9cm of center] (Tname) {Ty};
  \node [above right = -0.2cm and 0.3cm of center] (RTy) {$R_\text{Ty}$};
  \node [vertex, below right = 0.4cm and 0.9cm of center] (t1) {$t_1$};
  \node [vertex, below = 0.5cm of t1] (t2) {$t_2$};
  \draw [dashed] (1.2,-1.175) ellipse (0.4cm and 1cm);

  \draw[->] (u1) to[] (i1) ;
  \draw[->] (u1) to[] (i2) ;
  \draw[->] (u2) to[] (i1) ;
  \draw[->] (u2) to[] (i3) ;
  \draw[->] (i1) to[] (t1) ;
  \draw[->] (i2) to[] (t1) ;
  \draw[->] (i3) to[] (t2) ;

\end{tikzpicture}
\begin{tikzpicture}[{
  > = {Stealth [inset = 0pt, length = 10pt, angle' = 30, round]},
  vertexset/.style = {minimum width = 0.5cm, minimum height = 0.5cm, inner sep = 0pt},
  vertex/.style = {minimum width = 0.25cm, minimum height = 0.5cm, inner sep = 0pt, draw, circle},
  marked/.style = {draw = blue, color = blue, line width = (4*#1), text = blue},
  edge/.style={->}
}]
\scriptsize
  \node (center) {};
  \node [vertex, below left = 0.4cm and 1.4cm of center,color=blue] (u1) {$1/2$};
  \node [vertex, below = 0.5cm of u1,color=blue] (u2) {$1/2$};
  \draw [dashed] (-1.7,-1.175) ellipse (0.4cm and 1cm);
  \node [vertex, below left = 0.0cm and 0cm of center,color=blue] (i1) {$2/4$};
  \node [vertex, below = 0.5cm of i1,color=blue] (i2) {$1/4$};
  \node [vertex, below = 0.5cm of i2,color=blue] (i3) {$1/4$};
  \draw [dashed] (-0.29,-1.3) ellipse (0.5cm and 1.5cm);
  \node [vertex, below right = 0.4cm and 0.9cm of center,color=blue] (t1) {$3/4$};
  \node [vertex, below = 0.5cm of t1,color=blue] (t2) {$1/4$};
  \draw [dashed] (1.2,-1.175) ellipse (0.4cm and 1cm);

  \draw[->,color=blue] (u1) to[] (i1) ;
  \draw[->,color=blue] (u1) to[] (i2) ;
  \draw[->,color=blue] (u2) to[] (i1) ;
  \draw[->,color=blue] (u2) to[] (i3) ;
  \draw[->,color=blue] (i1) to[] (t1) ;
  \draw[->,color=blue] (i2) to[] (t1) ;
  \draw[->,color=blue] (i3) to[] (t2) ;

  \node [below right = 2.9cm and -1.25cm of center] (hin) {$p_\Pi = (3/4,1/4)$};

\end{tikzpicture}
\begin{tikzpicture}[{
  > = {Stealth [inset = 0pt, length = 10pt, angle' = 30, round]},
  vertexset/.style = {minimum width = 0.5cm, minimum height = 0.5cm, inner sep = 0pt},
  vertex/.style = {minimum width = 0.25cm, minimum height = 0.5cm, inner sep = 0pt, draw, circle},
  marked/.style = {draw = blue, color = blue, line width = (4*#1), text = blue},
  edge/.style={->}
}]
\scriptsize
  \node (center) {};
  \node [vertex, below left = 0.4cm and 1.4cm of center] (u1) {};
  \node [vertex, below = 0.5cm of u1,color=blue] (u2) {$1$};
  \draw [dashed] (-1.7,-1.175) ellipse (0.4cm and 1cm);
  \node [vertex, below left = 0.0cm and 0cm of center,color=blue] (i1) {$1/2$};
  \node [vertex, below = 0.5cm of i1,color=blue] (i2) {$0$};
  \node [vertex, below = 0.5cm of i2,color=blue] (i3) {$1/2$};
  \draw [dashed] (-0.29,-1.3) ellipse (0.5cm and 1.5cm);
  \node [vertex, below right = 0.4cm and 0.9cm of center,color=blue] (t1) {$1/2$};
  \node [vertex, below = 0.5cm of t1,color=blue] (t2) {$1/2$};
  \draw [dashed] (1.2,-1.175) ellipse (0.4cm and 1cm);

  \draw[->] (u1) to[] (i1) ;
  \draw[->] (u1) to[] (i2) ;
  \draw[->,color=blue] (u2) to[] (i1) ;
  \draw[->,color=blue] (u2) to[] (i3) ;
  \draw[->,color=blue] (i1) to[] (t1) ;
  \draw[->,color=blue] (i2) to[] (t1) ;
  \draw[->,color=blue] (i3) to[] (t2) ;

  \node [below right = 2.9cm and -1.75cm of center] (hin) {$p_\Pi(u_2) = (1/2,1/2)$};

\end{tikzpicture}
}
\caption{\label{fig:example_diagrams} Example of a HIN (left) having a schema as in Figure~\ref{fig:smallest}, with 2 users rating 3 items classified into 2 types, and computation of $p_\Pi$ (middle) and $p_\Pi(u_2)$ (right) for meta-path $\Pi = R_{rates} R_{Ty} $.
}
\end{figure}

Given a HIN $\HIN=(\Graph,\NodeLabels,\EdgeLabels,\NodeMap,\EdgeMap)$ let us consider
a random walk along a meta-path $\Pi$ starting in $S\in\NodeLabels$ and ending in $T\in\NodeLabels$.
Let us also consider the random variable $X$, 
the node of $T$ where the random walk along $\Pi$ ends.
We are interested in the probability mass function (PMF) over the nodes of $T$, denoting $p_{\Pi}$ the PMF $P(X=t\in T)$ when the walk started randomly at any node of $S$.
Similarly, let us denote by $p_{\Pi}(s)$ the PMF $P(X=t\in T|s\in S)$, when the walk started at the node $s\in S$.
To illustrate this better, let us consider a HIN  with 2 users rating 3 items classified in 2 types, corresponding to the schema in Figure~\ref{fig:smallest}.
In Figure~\ref{fig:example_diagrams}, we illustrate the computation of PMFs for such random walks on the simple example presented in Figure~\ref{fig:smallest}, following the meta-path $\Pi = R_{rates} R_{Ty} $: $p_\Pi$ when starting randomly in a set of users $U=\{u_1,u_2\}$, and $p_\Pi(u_2)$ when starting from user $u_2$.

Different diversity measures can then be computed from these PMFs (\eg Shannon entropy, Gini coefficient). 
We use \textit{perplexity}, denoted by $\diver$ and computed for a PMF $p$ as
$\diver(p)=2^{E(p)}$,  with  $E(p)=-\sum\nolimits^{|T|}_{i=1}p_i\log_2 (p_i)$, where
$ E(p) $ is the Shannon entropy of the PMF.
We select perplexity because of its intuitive interpretation: it measures the degree of unexpectedness of a random variable.
Perplexity accounts for \textit{variety} (number of possibilities available) and \textit{balance} (how probability is distributed among those possibilities)~\cite{stirling2007general}.
Interestingly, the perplexity of a PMF $p=(p_1, p_2, \ldots)$ can be interpreted as the number of elements of a uniform PMF that has the same Shannon entropy as $p$.
For instance,
$p= \left( \frac{1}{3},\frac{1}{3},\frac{1}{3} \right)$ has perplexity $\diver(p)=3$ (i.e., uniform over 3 elements) while
$p'= \left( \frac{6}{10},\frac{2}{10},\frac{1}{10},\frac{1}{10} \right)$ has perplexity $\diver(p')=2.97$, \ie $p'$ has the same entropy as a uniform distribution over 2.97 elements.
So $p'$ is a distribution over four possibilities, but has lower diversity than $p$.

Using these notions, and following~\cite{ramaciotti2020measuring}, we define the two following diversity measures for a meta-path $\Pi$ of a HIN $\HIN$:

\begin{gather*}
\text{Mean Individual diversity: } \midiv{\Pi} = \left(\prod\limits_{s\in S} \diver(p_\Pi(s))\right)^{1/|S|},\\
\text{Collective diversity: } \coldiv{\Pi} = \diver(p_\Pi),
\end{gather*}

\noindent with $\diver$ being the perplexity.

The Mean Individual (MI) diversity along a meta-path $\Pi$ is the geometric mean of the perplexities of the PMFs corresponding to random walks starting on each element $s\in S$ and ending in $T$.
It can thus be interpreted as the diversity in terms of elements of $T$ available on average for one element of~$S$.
Collective (Col) Diversity along meta-path $\Pi$ is the perplexity of the PMF corresponding to a single random walk starting randomly on an element $s\in S$ and ending in $T$.
It may be interpreted as the diversity in terms of elements of $T$ available to the entire group of elements of~$S$.
In the example of Figure~\ref{fig:example_diagrams}, the collective diversity of types available to users is $\coldiv{\Pi}=\diver(3/4,1/4) = 1.7547$, and the mean individual diversity is 
$$
\midiv{\Pi}=\left(\diver\right.\underbrace{(1,0)}_{p_\Pi(u_1)}\cdot\, \diver\underbrace{(1/2,1/2)}_{p_\Pi(u_2)}\left. \right)^{\frac{1}{2}}=\left(1 \cdot 2 \right)^{\frac{1}{2}} = \sqrt{2}.
$$

\subsection{Recommendations as Processes on HINs}
\label{sec:rs_and_hins}

A recommendation process consists in proposing items to a user that he or she has not selected before. 
In a HIN framework, it amounts to creating new edges:

\begin{definition}[Recommendation on a HIN]
A recommendation process on a HIN $\HIN=(\Graph,\NodeLabels,\EdgeLabels,\NodeMap,\EdgeMap)$ is a process $F(\HIN)$ that produces a new collection of edges
$F: \HIN \mapsto \EdgeSet_{rec}\in \NodeSet\times\NodeSet$.
The edge set $\EdgeSet_{rec}$ is composed of links between users and their recommended items.
\end{definition}

A HIN can then be updated with a new relation $R_{rec}$ such that $\EdgeMap(e)=R_{rec}$ for $e\in\EdgeSet_{rec}$, so that recommended edges can be used in meta-paths.
In the following section, we are interested in including recommendation edges in meta-paths to measure the diversity related to these paths, as described in Subsection~\ref{sec:random_walks}.
We may then measure how {\em diverse} the recommendation process is.

\subsection{A HIN Graph Spreading Recommender System}
\label{sec:graph_spreading}

In order to produce recommendations that can include different parts of a HIN in a controlled and interpretable fashion, we consider a recommendation procedure well-known to the RS community (cf.~\cite{sun2011pathsim}).
This method belongs to a larger family of RS based on random walks (\eg~\cite{nandanwar2018fusing,jiang2019degenerate}).

Precisely, it consists in computing the probability produced by random walks along a meta-path, starting on a single user and ending on the set of items.
Then, we simply recommend the items with the highest probability that have not been previously chosen.
This allows us to measure the effect of different parts of a HIN in an interpretable way.
Indeed, we can consider different meta-paths that include different meta-data (e.g., book genre, user location), and produce different probabilities that can be added with different coefficients to modulate their relative importance in recommendations.
For instance, we may give a high weight to the meta-path related to demographic information about the users and a light weight to information related to items typologies, or vice versa.
Let us formalize this type of recommendation by considering $K$ different meta-paths $ \Pi_k $ (with $k = 1,\ldots,K$) starting on $S\in\NodeLabels$
and ending in $T\in\NodeLabels$.
Let us consider the parameters $\alpha_1,\ldots,\alpha_K$, such that
$\sum \alpha_k = 1$ and $0\leq\alpha_k\leq 1$.
For every element $s\in S$ (typically the  user group), we compute a score $\mathcal{S}$ as a PMF over $T$ (typically the item group):
\begin{equation}
\mathcal{S}(t|s) = \sum\limits_{k=1}^{K} \alpha_k p_{\Pi_k}(s), \text{ for }t\in T.
\label{eq:sgrs}
\end{equation}
Using the score $\mathcal{S}(t|s)$ we recommend the top ranking elements of $T$ for $s\in S$ (among those not already selected).
This RS shares some principles with the one detailed by Zhou et al.~\cite{zhou2010solving}, in which mass (called \textit{resource} in~\cite{zhou2010solving}) is assigned to items and then distributed through edges to find relevant items according to the final resource distribution.

It should be noticed that this RS is not new and our point is not to argue that it performs better than other methods.
However, we show in the next section that its accuracy is comparable to that of popular methods.
Its main interest here is that it allows us to include different parts of a HIN in the recommendation process to measure separately the effects on accuracy and diversity of these parts.
 

\section{Experiments}
\label{sec:experiments}

Here we present the data used, the protocols for our experiments\footnote{The codes and data used for these experiments are available at \url{https://github.com/pedroramaciotti/HINPy}}, and their results to answer our research questions.

\subsection{Datasets and Meta-Path Diversities of HIN}
\label{subsec:datasets}

\begin{figure}[H]
\centering
\resizebox{0.8\columnwidth}{!}{%
\begin{tabular}{|c|r|r|r|r|r|r|}
\hline
\multicolumn{6}{|c|}{MovieLens 100K (ML100K)}\\
\hline
Relation & \makecell[c]{Start Obj.\\ Group} & Size & \makecell[c]{End Obj.\\ Group} & Size & Links \\
\hline
$R_{\text{Ty}}$ & movie & 1664 & type & 19 & 2863  \\
$R_{\text{Ye}}$ & movie & 1664 & release & 72 & 1664  \\
$R_{\text{rates}}$ & user & 943 & movie & 1664 & 99965  \\
$R_{\text{Oc}}$ & user & 943 & occupation & 21 & 943  \\
$R_{\text{Ag}}$ & user & 943 & age group & 61 & 943  \\
$R_{\text{Ge}}$ & user & 943 & gender & 2 & 943  \\
$R_{\text{Lo}}$ & user & 943 & location & 795 & 943  \\
$R_{\text{likes}}$ & user & 943 & movie & 1664 & 82495  \\
\hline
\end{tabular}
}

\vspace{0.2cm}

\begin{tikzpicture}[{
  > = {Stealth [inset = 0pt, length = 5pt, angle' = 30, round]},
  vertexset/.style = {minimum width = 0.5cm, minimum height = 0.5cm, inner sep = 0pt},
  vertex/.style = {minimum width = 0.5cm, minimum height = 0.5cm, inner sep = 0pt, draw, circle},
  marked/.style = {draw = blue, color = blue, line width = (4*#1), text = blue},
  edge/.style={->}
}]
\tiny
  \node [vertex] (u) {U};
  \node [vertex, right = 1cm of u] (i) {I};
  \node [vertex, above right = 0.75cm of i] (Ty) {Ty};
  \node [vertex, below right = 0.75cm of i] (Ye) {Ye};

  \node [vertex, above = 0.75cm of u] (Oc) {Oc};
  \node [vertex, above left = 0.5cm and 0.866025cm of u] (Ge)  {Ge};
  \node [vertex, below left = 0.5cm and 0.866025cm of u] (Lo)  {Lo};
  \node [vertex, below = 0.75cm of u] (Ag)  {Ag};

  \draw[->] (u) to[out=45, in=135, edge node={node [midway,above] {$R_{\text{rates}}$}}] (i) ;
  \draw[dashed, ->] (u) to[out=0, in=180, edge node={node [midway,above] {$R_{\text{likes}}$}}] (i) ;
  \draw[dashed,->] (u) to[out=-45, in=225, edge node={node [midway,above] {$R_{\text{rec}}$}}] (i) ;
  \draw[->] (i) to[out=45, in=225, edge node={node [midway,right] {$R_{\text{Ty}}$}}] (Ty);
  \draw[->] (i) to[out=-45, in=135, edge node={node [midway,right] {$R_{\text{Ye}}$}}] (Ye);
  \draw[->] (u) to[out=90,  in=-90, edge node={node [midway,left] {$R_{\text{Oc}}$}}] (Oc);
  \draw[->] (u) to[out=145, in=-35, edge node={node [midway,left] {$R_{\text{Ge}}$}}] (Ge);
  \draw[->] (u) to[out=215, in=40, edge node={node [midway,right] {$R_{\text{Lo}}$}}] (Lo);
  \draw[->] (u) to[out=270, in=90, edge node={node [midway,right] {$R_{\text{Ag}}$}}] (Ag);

  \node [below right= 0cm and 2.6cm of Oc,draw,align=left] (legend){
  \begin{tabular}{ll}
  Ty: Type of film    & $R_{\text{Ty}}$: Is of type            \\
  Ye: Year of release & $R_{\text{Ye}}$: Was released on year  \\
  Oc: Occupation      & $R_{\text{Oc}}$: Has occupation        \\
  Ge: Gender          & $R_{\text{Ge}}$: Is of gender          \\
  Lo: Location        & $R_{\text{Lo}}$: Is in location        \\
  Ag: Age             & $R_{\text{Ag}}$: Is of age             \\
                      & $R_{\text{rates}}$: Rates    \\
  U: User             & $R_{\text{likes}}$: Likes    \\
  I: Item             & $R_{\text{rec}}$: Is recommended    \\
  \end{tabular}
  };

\end{tikzpicture}

\vspace{0.4cm}

\resizebox{0.8\columnwidth}{!}{%
\begin{small}
\begin{tabular}{|c|r|r|r|r|r|r|}
\hline
\multicolumn{6}{|c|}{Douban Movie (DM)}\\
\hline
Relation & \makecell[c]{Start Obj.\\ Group} & Size & \makecell[c]{End Obj.\\ Group} & Size & Links  \\
\hline
$R_{\text{Ac}}$ & movie & 6971 & actor & 3004 & 15584  \\
$R_{\text{Di}}$ & movie & 6971 & director & 789 & 3314  \\
$R_{\text{Ty}}$ & movie & 6971 & type & 36 & 15598  \\
$R_{\text{Gr}}$ & user & 3022 & usergroup & 2269 & 7054  \\
$R_{\text{Lo}}$ & user & 3022 & location & 244 & 2491  \\
$R_{\text{rates}}$ & user & 3022 & movie & 6971 & 195493  \\
$R_{\text{Fr}}$ & user & 3022 & user & 3022 & 779  \\
$R_{\text{likes}}$ & user & 3022 & movie & 6971 & 182069  \\
\hline
\end{tabular}
\end{small}
}

\vspace{0.2cm}

\begin{tikzpicture}[{
  > = {Stealth [inset = 0pt, length = 5pt, angle' = 30, round]},
  vertexset/.style = {minimum width = 0.5cm, minimum height = 0.5cm, inner sep = 0pt},
  vertex/.style = {minimum width = 0.5cm, minimum height = 0.5cm, inner sep = 0pt, draw, circle},
  marked/.style = {draw = blue, color = blue, line width = (4*#1), text = blue},
  edge/.style={->}
}]
\tiny
  \node [vertex] (u) {U};
  \node [vertex, right = 1cm of u] (i) {I};
  \node [vertex, above = 0.75cm of i] (Ac) {Ac};
  \node [vertex, right = 0.75cm of i] (Di) {Di};
  \node [vertex, below = 0.75cm of i] (Ty) {Ty};

  \node [vertex, above = 0.75cm of u] (Lo) {Lo};
  \node [vertex, left = 0.75cm of u] (Gr)  {Gr};

  \draw[->] (u) to[out=45, in=135, edge node={node [midway,above] {$R_{\text{rates}}$}}] (i) ;
  \draw[dashed,->] (u) to[out=0, in=180, edge node={node [midway,above] {$R_{\text{likes}}$}}] (i) ;
  \draw[dashed,->] (u) to[out=-45, in=225, edge node={node [midway,above] {$R_{\text{rec}}$}}] (i) ;
  \draw[->] (i) to[out=90, in=-90, edge node={node [midway,right] {$R_{\text{Ac}}$}}] (Ac);
  \draw[->] (i) to[out=0,  in=180, edge node={node [midway,above] {$R_{\text{Di}}$}}] (Di);
  \draw[->] (i) to[out=-90, in=90, edge node={node [midway,right] {$R_{\text{Ty}}$}}] (Ty);
  \draw[->] (u) to[out=90,  in=-90, edge node={node [midway,left] {$R_{\text{Lo}}$}}] (Lo);
  \draw[->] (u) to[out=180, in=0, edge node={node [midway,above] {$R_{\text{Gr}}$}}] (Gr);
  \draw[->] (u) to[out=225, in=-90, min distance=10mm,edge node={node [midway,left] {$R_{\text{Fr}}$}}] (u);

  \node [below right= 0cm and 1.4cm of Ac,draw,align=left] (legend){
  \begin{tabular}{ll}
  Ty: Type of film    & $R_{\text{Ty}}$: Is of type    \\
  Ac: Actor           & $R_{\text{Ac}}$: Stars    \\
  Di: Director        & $R_{\text{Di}}$: Is directed by    \\
  Lo: Location        & $R_{\text{Lo}}$: Is in location    \\
  Gr: Group           & $R_{\text{Gr}}$: Belongs to    \\
                      & $R_{\text{Fr}}$: Is friends with    \\
  U: User             & $R_{\text{rates}}$: Rates    \\
  I: Item             & $R_{\text{likes}}$: Likes    \\
                      & $R_{\text{rec}}$: Is recommended    \\
  \end{tabular}
  };
 
\end{tikzpicture}

\caption{Summary and HIN schema of the MovieLens 100K (top) and Douban Movie (bottom) datasets.}
\label{fig:datasets_schemas_and_table}
\end{figure}
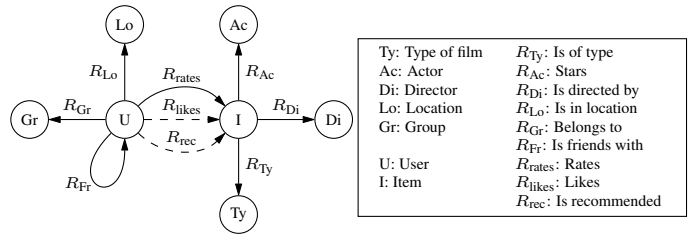

We use two datasets in our experiments, the well-documented MovieLens 100K dataset (ML100K)~\cite{harper2016movielens} and the Douban Movie (DM) dataset~\cite{Zafarani+Liu:2009}. 
The latter has been used in HIN-based RS studies, as it presents a rich HIN structure~\cite{shi2015semantic}.
We detail in Figure~\ref{fig:datasets_schemas_and_table} the characteristics of these datasets: groups of entities, their sizes, and the relations that join them.
We also show the corresponding HIN schemas with their object groups and link groups (relations).
Both datasets contain information on the rating (from 1 to 5) given to films by users, and the fact that a user has rated a movie is represented by the relation $R_{\text{rating}}$. 
%
We consider that a user likes a film if he or she rated it 3 or more (for both datasets), and we refer to the corresponding relation as $R_{\text{likes}}$.
Figure~\ref{fig:datasets_schemas_and_table} also shows the liking and recommendation relations with dashed lines, $R_{\text{likes}}$ and $R_{\text{rec}}$, the latter indicating that a RS may propose a recommendation of an item to a user.

\label{subsec:exploring_diversities} 


Now, we can illustrate the use of the diversity measures on HINs defined in Section~\ref{sec:preliminaries} with the ML100K and DM datasets.
While diversity measures from the literature essentially rely on how proposed or selected items are distributed into types,
the diversity measures described in Section~\ref{sec:random_walks} allow to consider a larger spectrum of diversity notions.
Let us consider the example of a diversity computed on meta-path $\Pi_1 = R_{\text{rec}}R_{\text{Ty}}$.
It is the diversity of types of movies in a recommendation list, consequently it captures the same intuition as the \textit{Intra-List Similarity} does in~\cite{ziegler2005improving}.
However, the meta-path based diversity allows us to express other forms of diversities.
For instance, the meta-path $\Pi_2 = R_{\text{rec}}R^{-1}_{\text{likes}}R_{\text{likes}}R_{\text{Ty}}$ relates to the diversity of types of movies liked by users who also liked movies that they were recommended.

\begin{figure}[!h]
  \begin{subfigure}[c]{0.49\textwidth} 
        \centering 
 \includegraphics[width=\columnwidth]{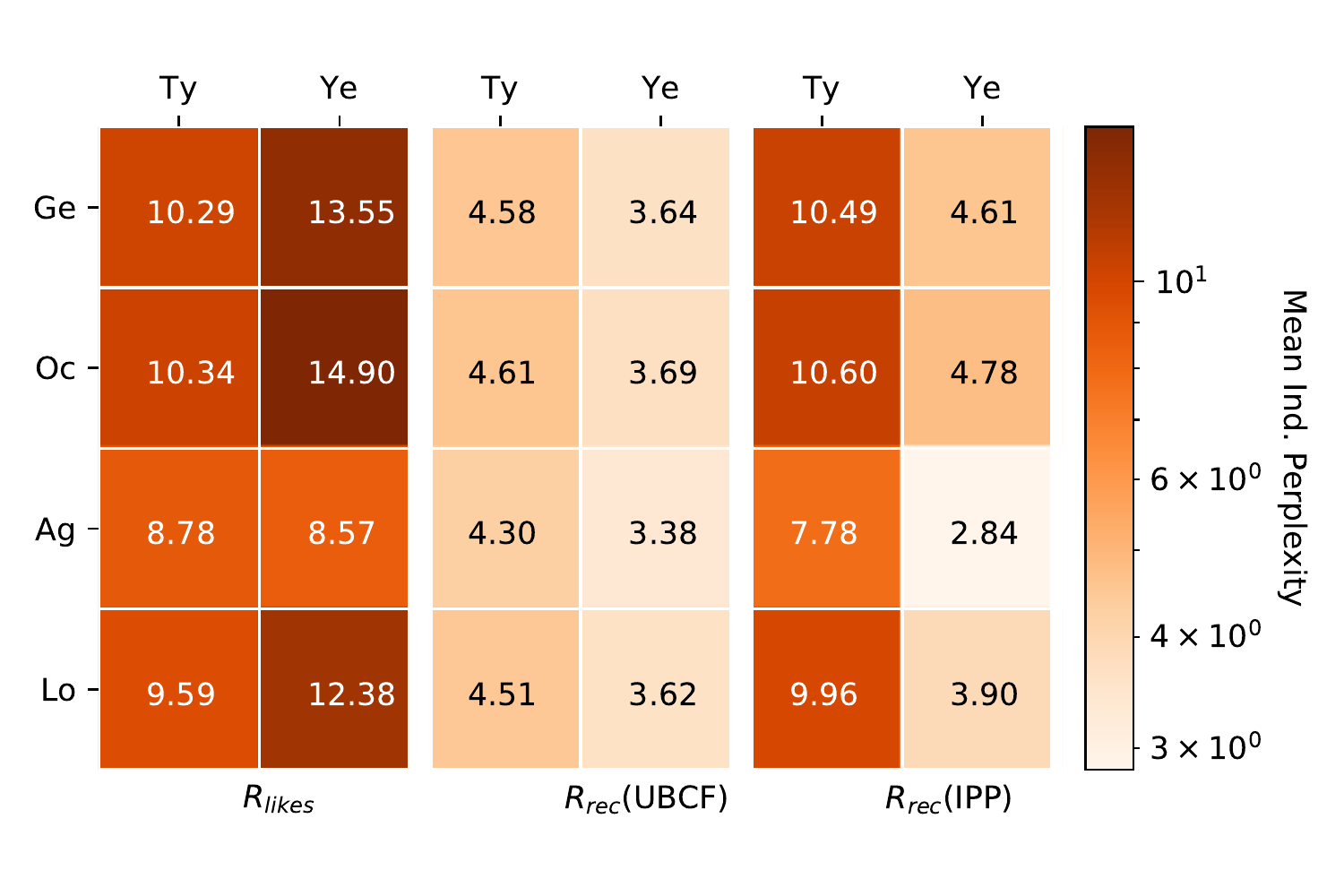}
        \caption{ML100K}\label{fig:orchid}
    \end{subfigure}
    ~ 
    \begin{subfigure}[c]{0.49\textwidth}
        \centering 
 \includegraphics[width=\columnwidth]{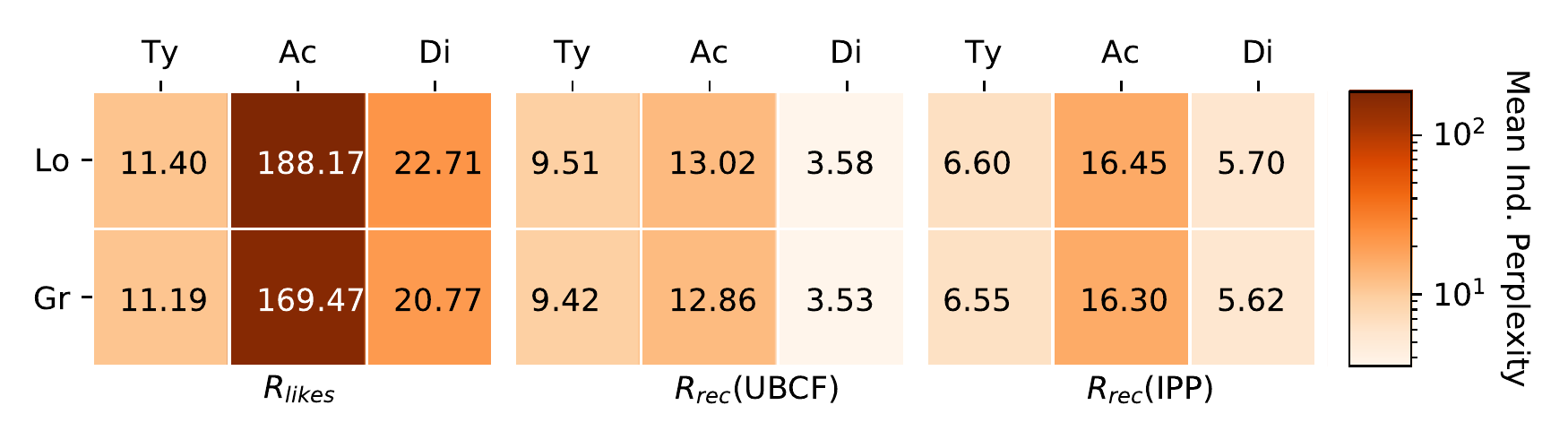}
        \caption{Douban}\label{fig:calla}
    \end{subfigure}

\caption{Mean individual diversities of different meta-paths $\Pi=R^{-1}_{S} R_{X} R_{\text{T}} $ for ML100K (a) and DM (b) datasets. 
$S$ are indicated in rows, $T$ in columns, and $X$ at the bottom for each group of diversities (5 items recommended per user).}
\label{fig:mosaic_reco_div_paths}
\end{figure}

In Figure~\ref{fig:mosaic_reco_div_paths}, we report experimental diversity measures on both datasets.
We consider meta-paths of length 3 starting in $S$ and ending in $T$ (with $S,T\in\NodeLabels$), which are node groups containing  semantic information about users and items respectively, so $\Pi = R^{-1}_{S} R_{X} R_{\text{T}}$.
Here $X$ stands for \textit{rating}, \textit{liked} or \textit{recommended}.
Then, we compute the Mean Individual diversity on these paths: $ \midiv{\Pi} $.
Regarding the recommendation links, we use two standard RSs in these experiments, each recommending a list of 5 items to the users: \textit{User-Based Collaborative Filtering}~(UBCF) and \emph{Implicit Pure Popularity}~(IPP), IPP recommends to the users the top-5 most liked movies, not already rated by the user.
The figure reads as follows: it reports Mean Individual diversities for all combinations of node groups $S$ and $T$ and $R_X$ is either $R_{likes}$, $R_{rec}(UBCF)$ or  $R_{rec}(IPP)$.
For instance, $\midiv{R^{-1}_{Lo} R_{rec}(UBCF) R_{\text{Ty}} }$ is the Mean Individual diversity of movie types recommended to users using UBCF, averaged over locations.

The measurements of these diversities provide insights about the structure of the HIN and about the RS.
First, notice that most $ \midiv{\Pi} $ using \textit{recommendations} are lower than those using \textit{likes}. 
This is only a consequence of the fact that recommended lists are of length 5, while the average user likes larger numbers of items in these datasets.
Similarly, the fact that $\midiv{R^{-1}_{S} R_{X} R_{\text{Ac}}}$ is much larger than $\midiv{R^{-1}_{S} R_{X} R_{\text{Ty}}}$ or $\midiv{R^{-1}_{S} R_{X} R_{\text{Di}}}$ on DM dataset reflects the fact that movies have more actors than directors or types/genres.
Other phenomena are revealed by these diversity measures.
For example, the diversity of release year of films by age groups (\ie of $R^{-1}_\text{Ag}R_\text{likes}R_\text{Ye}$) in the ML100K dataset is low.
This could point to the fact that users from a given age group tend to choose films from a narrow period of time, however testing this hypothesis falls outside the scope of this article.
Note also that these diversity measures allow us to compare diversities from one RS to another.
For example UBCF produces recommendations that are consistently less diverse  than those produced using IPP on the ML100K dataset.


\subsection{Effect of the HIN Structure on Accuracy and Diversity of Recommendations}
\label{subsec:structure_accuracy_diversity} 

Having established and illustrated the use of meta-path diversities, we now focus on the effect that the structure of HINs has on accuracy and diversity of recommendations.
For this purpose, we use the RS described in Section~\ref{sec:graph_spreading}.
It allows us to include in the recommendation process different parts of a HIN in a controlled way.
Moreover, its underlying mechanic, based on meta-path constrained random walks, is common to several techniques of HIN-based recommendation (e.g., \cite{sun2011pathsim,yu2013recommendation,shi2015semantic}).
Note that our goal here is to explore the consequences on accuracy and diversity of integrating different parts of a HIN, rather than optimizing this RS to obtain the best possible performance.

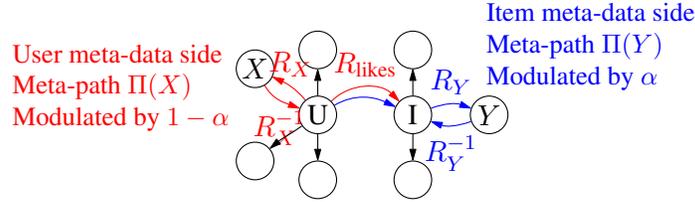
\begin{figure}[!h]
\begin{center}
\begin{tikzpicture}[{
  > = {Stealth [inset = 0pt, length = 5pt, angle' = 30, round]},
  vertexset/.style = {minimum width = 0.5cm, minimum height = 0.5cm, inner sep = 0pt},
  vertex/.style = {minimum width = 0.5cm, minimum height = 0.5cm, inner sep = 0pt, draw, circle},
  marked/.style = {draw = blue, color = blue, line width = (4*#1), text = blue},
  edge/.style={->}
}]
\small
  \node [vertex] (u) {U};
  \node [vertex, right = 0.75cm of u] (i) {I};
  \node [vertex, above = 0.35cm of i] (Ty) {};
  \node [vertex, below = 0.35cm of i] (Ye) {};
  \node [vertex, right = 0.5cm of i] (New) {$Y$};

  \node [vertex, above = 0.35cm of u] (Oc) {};
  \node [vertex, above left = 0.25cm and 0.47cm of u] (Ge)  {$X$};
  \node [vertex, below left = 0.25cm and 0.47cm of u] (Lo)  {};
  \node [vertex, below = 0.35cm of u] (Ag)  {};
  \node [below left = -0.45cm and 0.65cm of Oc,red] (utext)  {\footnotesize \begin{tabular}{l} User meta-data side \\ Meta-path $\Pi(X)$\\Modulated by $1-\alpha$\end{tabular}};
  \node [above right = -0.9cm and 0.45cm of Ty,blue] (utext)  {\footnotesize \begin{tabular}{l} Item meta-data side \\ Meta-path $\Pi(Y)$\\Modulated by $\alpha$\end{tabular}};

  \draw[->,red] (u) to[out=45, in=135, edge node={node [midway,above] {$R_\text{likes}$}}] (i) ;
  \draw[->,blue] (u) to[out=30, in=155,blue, edge node={node [midway,above] {}}] (i) ;
  \draw[->] (i) to[out=90, in=-90, edge node={node [midway,right] {}}] (Ty);
  \draw[->] (i) to[out=-90, in=90, edge node={node [midway,right] {}}] (Ye);
  \draw[->] (u) to[out=90,  in=-90, edge node={node [midway,left] {}}] (Oc);
  \draw[->,red] (u) to[out=130, in=-25, red, edge node={node [midway,above] {$R_X$}}] (Ge);
  \draw[->,red] (Ge) to[out=-60, in=160, red, edge node={node [midway,below] {$R^{-1}_X$}}] (u);
  \draw[->] (u) to[out=215, in=40, edge node={node [midway,right] {}}] (Lo);
  \draw[->] (u) to[out=270, in=90, edge node={node [midway,right] {}}] (Ag);

  \draw[->,blue] (i) to[out=20, in=160,blue, edge node={node [midway,above] {$R_Y$}}] (New);
  \draw[->,blue] (New) to[out=-160, in=-20,blue, edge node={node [midway,below] {$R^{-1}_Y$}}] (i);

\end{tikzpicture}
\end{center}
\caption{Schematic illustration of the recommendation procedure used in the experiments, where parameter $\alpha$ modulates two meta-paths, $\Pi(X)$ and $\Pi(Y)$, containing user and item meta-data.}
\label{fig:schema_mixed_paths}
\end{figure}

Many choices are available, as we can mix all kinds of meta-paths to produce a recommendation.
However, we choose to consider a RS that mixes only two meta-paths for each dataset: one denoted $\Pi(X)$ which includes a {\em user-content} object group (\eg the location of a user), and one $\Pi(Y)$ including an {\em item-content} object group (\eg the genre of a movie).
By tuning the balance between these two meta-paths, we can control what kind of information is taken into account to produce a recommendation.
Practically, Equation~(\ref{eq:sgrs}) here takes the form:
 $$
\mathcal{S}(t|s)=(1-\alpha) \left(p_{\Pi(X)}(s)\right)(t) +  \alpha\left(p_{\Pi(Y)}(s)\right)(t)
 $$
$s\in\Users$ and $t\in\Items$ and  parameter $\alpha\in[0,1]$ tunes the balance between $\Pi(X)$ and $\Pi(Y)$.
So, $\alpha=1$ means that only the item-content is used in computing recommendations and when $\alpha=0$ only the user-content is used.
Precisely, the user-content path $\Pi(X)$ is $R_{X}R^{-1}_X R_\text{likes}$ where $X$ stands for a user-content group, and the item-content path $\Pi(Y)$ is $R_\text{likes}R_{Y}R^{-1}_Y $ where $Y$ stands for an item-content group.
Figure~\ref{fig:schema_mixed_paths} illustrates the meta-paths selected on the HIN schema.

We use  $\alpha\in\{1.0,0.8,0.6,0.4,0.2,0.0\}$ to show the effect of gradually moving from recommendations purely based on item meta-data ($\alpha=1$) to recommendations based purely on user meta-data ($\alpha=0$).
While there are several possible combinations of meta-paths (8 combinations for ML100K and 6 for DM), we only focus on a few of them (4 for each dataset) for ease of illustration and also because they suffice to show the effects that the HIN structure has on accuracy and diversity.

\subsubsection{Accuracy}
\label{subsubsec:acc}

\begin{figure*}[!h]
  \begin{subfigure}[c]{0.49\textwidth} 
        \centering 
\includegraphics[width=\columnwidth]{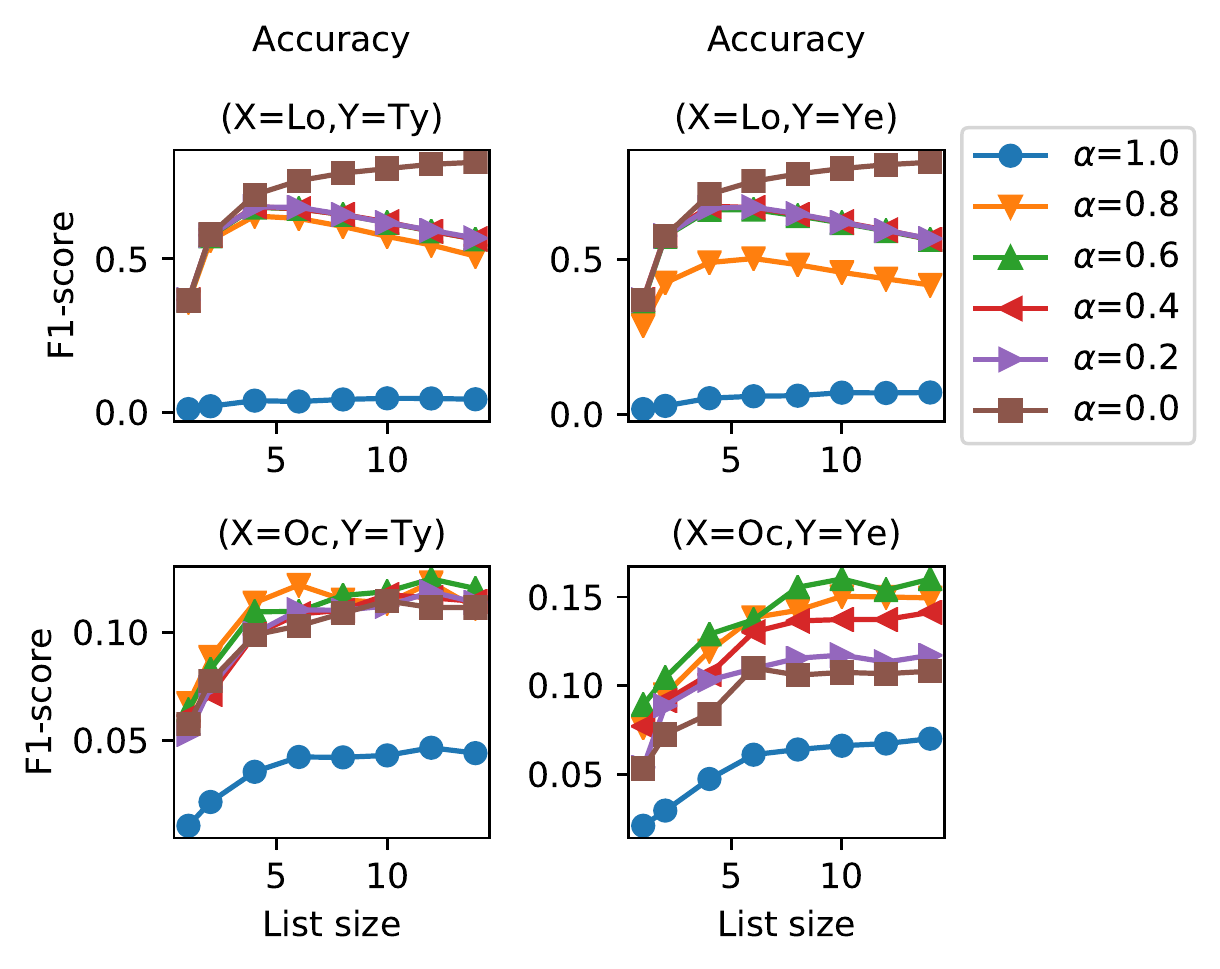}
\caption{MovieLens 100K}
\label{fig:accuracy_varying_ml100k}
    \end{subfigure}
    ~ 
    \begin{subfigure}[c]{0.49\textwidth}
        \centering 
\includegraphics[width=\columnwidth]{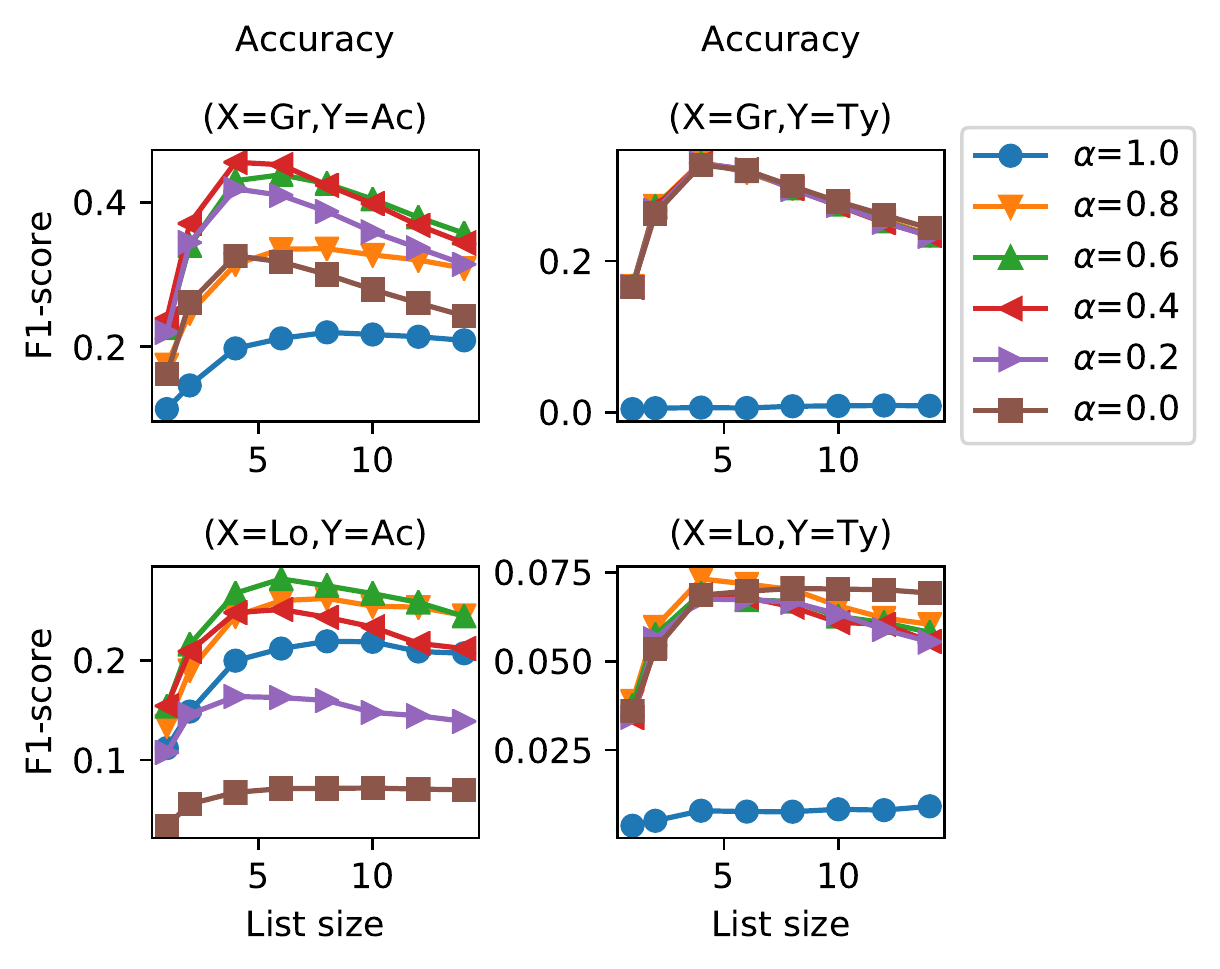}
\caption{Douban Movie}
\label{fig:accuracy_varying_douban_movie}
    \end{subfigure}
\caption{F1 score for different list sizes of recommended items and different values of $ \alpha $.
The meta-paths used are $\Pi(X)$ and $\Pi(Y)$, where $X$ is either Gender or Location (in rows) and $Y$ is either Actors or Type (in columns). }
\end{figure*}

Figures~\ref{fig:accuracy_varying_ml100k} and \ref{fig:accuracy_varying_douban_movie} show the F1-score of recommendations using the ML100K and DB datasets respectively.
%
%
For each experiment, we set a list size (the number of items included in a recommendation list proposed to each user).
For a given list size and $\alpha$, a random 10\% of user-item links are hidden.
We compute the items recommended to users (corresponding to the edges $R_\text{rec}$ in Figure~\ref{fig:datasets_schemas_and_table}) with our HIN-based RS and compare it to the test set to evaluate the F1-score of our prediction.
Each figure in the mosaic corresponds to a combination of meta-paths, with a given $X$ and a given $Y$.


While it is not our main objective to characterize the performance of this recommendation method, is it worth noticing that it may produce relevant recommendations: some combinations of meta-paths and $\alpha$ result in F1-scores comparable to other RS in different settings (see, \eg~\cite{hatami2014improving,shi2018heterogeneous}).
Note also that whatever the list size, the pure item-content recommendation is never the highest F1-score.
In other words, accuracy is, unsurprisingly, increased (up to a certain point) by the addition of user-content meta-data for both datasets.

On Douban Movie with a fixed list size, the best F1-scores are obtained with a mixture of item- and user-related information.
This is less true for ML100K, where user-related information seems often more valuable than item-related information.
We also notice that the range of F1-score values varies significantly depending on the balance between meta-paths: we see variations by up to a factor of 14 on ML100K (list of size 15 using user locations and movie types). 
This observation is consistent with the idea that while some meta-data brings relatively irrelevant information for recommendation (e.g. movie year of release in ML100K), other are much more valuable for the accuracy of the recommendation (e.g. user location in ML100K).

\subsubsection{Diversity}

\begin{figure*}[h]

\begin{center}
\begin{subfigure}[b]{0.7\textwidth} 
        \centering 
\includegraphics[width=1\linewidth]{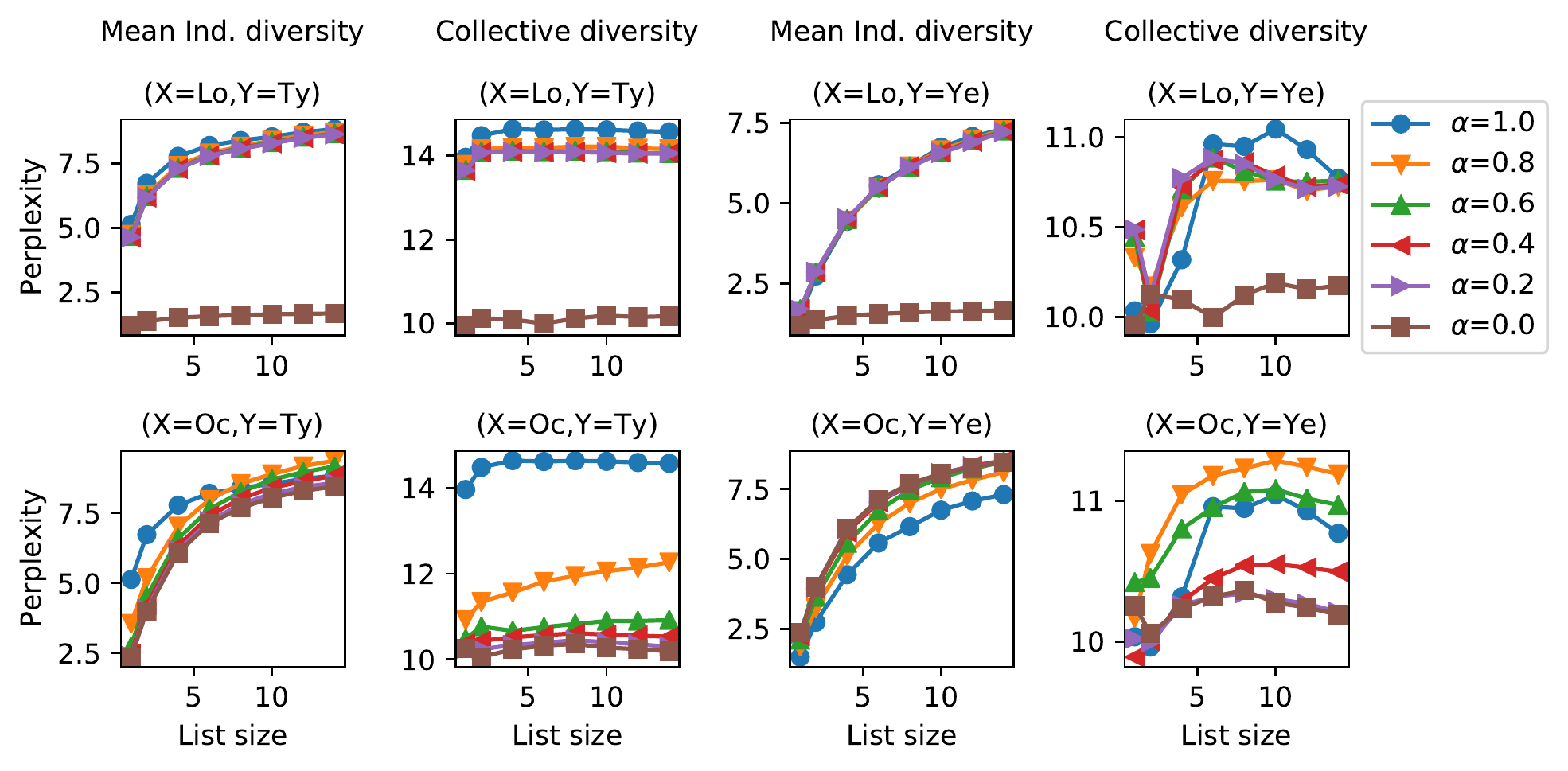}
\caption{MovieLens 100K}
\label{fig:diversity_varying_ml100k}
    \end{subfigure}
    
    \vspace*{3mm}
    
\begin{subfigure}[b]{0.7\textwidth}
        \centering 
\includegraphics[width=1\linewidth]{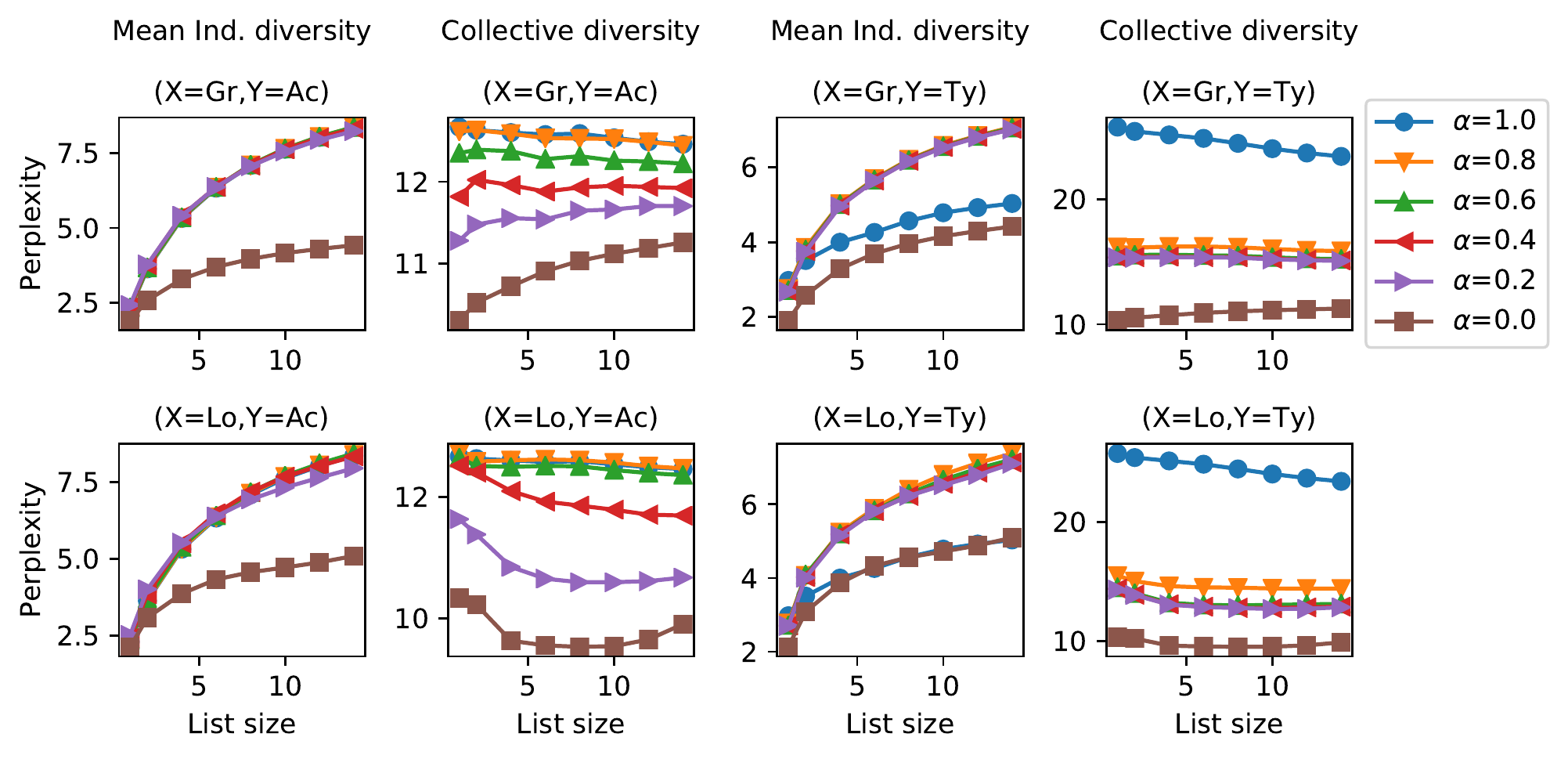}
\caption{Douban Movie}
\label{fig:diversity_varying_douban_movie}
    \end{subfigure}
\end{center}
    
    \caption{Collective and Mean Individual diversities in terms of types of items recommended to users, for different list sizes and recommendation meta-paths modulated by $\alpha$, for MovieLens 100K dataset (a), and Douban Movie dataset (b).}
\end{figure*}

For the same recommendation protocol, we now measure the Mean Individual and Collective diversities of the recommendation lists, in terms of types of items recommended.
In other words, we compute $ \midiv{\Pi} $ and $ \coldiv{\Pi} $ with the $\Pi=R_{\text{rec}}R_{\text{Ty}}$ meta-path.
The results are displayed in Figures~\ref{fig:diversity_varying_ml100k} and~\ref{fig:diversity_varying_douban_movie}, following the same display organization as figures in~\ref{subsubsec:acc}.

Figures~\ref{fig:diversity_varying_ml100k} and~\ref{fig:diversity_varying_douban_movie} highlight the interpretation associated with diversity measures based on meta-paths, and especially the contrast between Mean Individual diversity and Collective diversity.
For both datasets, Mean Individual diversity monotonically increases with the list size, which is expected: more items recommended to a user implies that the user may access more types.
In other words, \textit{variety} increases with the list size (although not necessarily \textit{balance}).
However, this tendency is limited as perplexity cannot be larger than the available number of types (19 in ML100K, 36 in DM).
While the Mean Individual diversity grows with the number of recommended items, Collective diversity may stagnate or even decrease.
For instance, we can read on DM dataset that Collective diversity decreases with the list size for all $ \alpha > 0$ when $ X = $ Gr (Groups of users) and $ Y =$ Ty (movie Types).
This result is subtle but important in our opinion.
It means that even if each user is individually recommended more diverse items, the whole group of users are recommended less diverse items.

Other trends can be isolated from these experiments, we report a few of them.
First, we observe that including user meta-data tends to improve accuracy, but also diminishes Collective diversity.
This is more obvious on ML100K dataset, where combinations which bring higher F1 scores, are often the ones which have the lowest diversities (both Collective and Mean Individual).
Also, including more user meta-data usually has little effect on Mean Individual diversity.
This observation is quite surprising as user meta-data can connect users of the HIN who potentially have very little in common (for example users of a same gender), so we would expect that these links give access to a large variety of items.
We think that these observations are certainly related to the HIN structure and we begin to address this question in the next section.


\subsection{Testing Semantic Relevance of Links Using the Configuration Model}
\label{subsec:acc_diversity_config_model}

The previous results raise some interesting questions.
Why does the inclusion of certain meta-paths changes the accuracy and the diversity more than others?
What is the property of user-content links that allows for more accurate recommendations?
A natural assumption is that they contain relevant semantic information about the origin of the users' preferences.

To test this hypothesis, we propose an experiment inspired by the field of social network analysis (SNA).
To identify the specificities of a real network in SNA, it is usual to compare its structure to a random graph model with a similar degree sequence, that is to say a graph in which nodes have the same number of connections as in the original graph, but distributed randomly.
This model is often referred to as the \textit{configuration model}~\cite{newman2018networks}.
A usual way to perform such a comparison consists in iteratively shuffling the links of the existing graph by exchanging their extremities.
Here, we propose to follow this process for the edges of a specific relation (link group) of the HIN.
To illustrate this idea, let's consider the example of the user-content links $R_\text{Lo}$: by shuffling only those links, we do not modify the number of individuals per location.
Thus we leave an essential property of the original HIN untouched, that is its degree sequence.
However, we empty this part of the HIN of its semantic meaning, because the location associated to a user is now random.

As for the protocol itself, we proceed as follows: we focus on a recommendation based on the mixing of two meta-paths, one using user-based information and another using item-based information, as we did previously.
We select the combinations that produced the most accurate recommendations: user location (Lo) and movie type (Ty) for ML100K, and user groups (Gr) and movie actors (Ac) for DM.
In addition to the recommendation produced on the original ML100K HIN (respectively DM HIN), which was reported in Section~\ref{subsec:structure_accuracy_diversity}, we make recommendations on several realizations of a HIN where $R_\text{Lo}$ (resp. $R_\text{Gr}$) have been randomly shuffled.
Then we compare the accuracy, Mean Individual and Collective diversities obtained on the real HIN to the values on the randomized HINs.
We better detail this procedure thereafter.

\begin{figure}[!h]
\centering
\resizebox{0.98\columnwidth}{!}{%
\begin{tabular}{|l|l|l|l|l|l|}
\hline
Dataset & \makecell[l]{User-content\\ meta-path} & 
\makecell[l]{Related \\user-content\\object group\\ (to be shuffled)} & 
\makecell[l]{Item-content\\ meta-path} & \makecell[l]{Related \\item-content\\object group} & \makecell[l]{Number of \\ randomizations} \\
\hline
ML100K & $R_{\text{Lo}}R^{-1}_{\text{Lo}}R_{\text{likes}}$ & Location (Lo) & $R_{\text{likes}}R_{\text{Ty}}R^{-1}_{\text{Ty}}$ & Type (Ty) & 100 \\
\hline
DM & $R_{\text{Gr}}R^{-1}_{\text{Gr}}R_{\text{likes}}$ & Usergroup (Gr) & $R_{\text{likes}}R_{\text{Ac}}R^{-1}_{\text{Ac}}$ & Actor (Ac) & 20 \\
\hline
\end{tabular}
}
\bigbreak
\includegraphics[width=0.98\columnwidth]{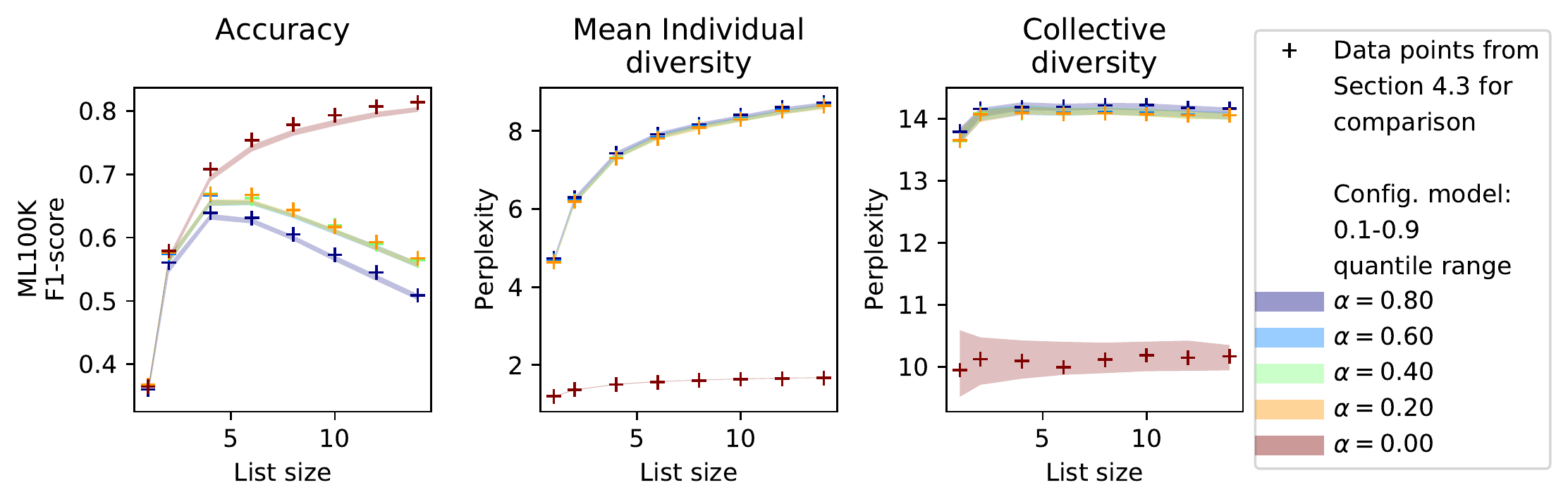}
\bigbreak
\includegraphics[width=0.98\columnwidth]{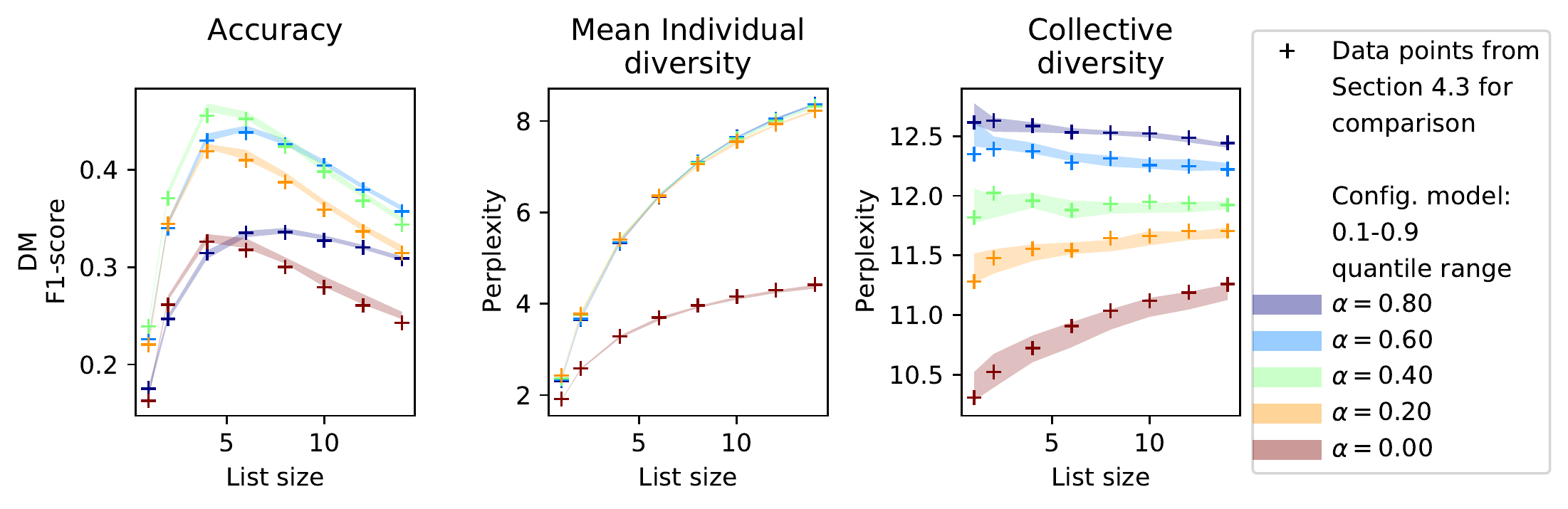}
\caption{Comparison between recommendations on randomized HINs compared to the original HIN:
0.1-0.9 quantile ranges for F1-score, Mean Individual, and Collective diversities for different list sizes, $\alpha$ values.}
\label{fig:detailed_config_model_ml100k}
\end{figure}

We produce 100 randomized versions of the HIN for the ML100K by shuffling the edges in the $R_\text{Lo}$ link group, and 20 randomized HIN for the DM dataset  by shuffling the edges in the $R_\text{Gr}$ link group. 
Then, for each list size and for $\alpha\in\{0.8,0.6,0.4,0.2,0.0\}$, we compute the 0.1--0.9 quantile range of the F1-score, Mean Individual, and Collective diversities.
Note that $\alpha=1$ produces recommendations using only item-related meta-paths and is of no interest as it does not include randomized parts.
Lower $\alpha$ values imply that randomized links from $R_\text{Lo}$ (ML100K) or $R_\text{Gr}$ (DM) are used in the recommendation process.
A value of $\alpha=0$ means that only the suffled link group $R_\text{Lo}$ (ML100K) or $R_\text{Gr}$ (DM) was used in computing recommendations.
The resulting quantile ranges, compared to the values obtained with the original HIN from Section~\ref{subsec:structure_accuracy_diversity} are shown in Figure~\ref{fig:detailed_config_model_ml100k}.

We observe in all cases that the recommendations on the original HIN and on its randomized versions are comparable in terms of accuracy and diversities.
These results are counter-intuitive: it seems that emptying important parts of the HIN of its semantic content does not have a significant impact on the recommendation.
These results are surprising as our recommendation process relies exclusively on the number of paths in the HIN connecting a user to an item.
Our results are of course limited to the selected parameters and datasets used and call for further investigation in other contexts.
Still, they tend to disprove the underlying hypothesis that the relevance of meta-path recommendations only comes from the semantic information that they contain.
They could indicate that the number of connections of a node (which is not affected by the randomization) is the most -- and even the only -- important information in the recommendation process based on the HIN structure.
This new hypothesis is not tested in this article as it demands a wider experimental setting than the one considered for the purposes stated in the introduction. 


\section{Conclusion} 

We used recently proposed diversity measures for the evaluation of recommendations on HINs.
They are based on the {\em perplexity} of PMFs resulting from random walks along meta-paths over a Heterogeneous Information Network.
While standard diversities measures usually focus on the distribution of the recommended items into types, we may measure diversity on an individual and collective level, revealing that these two approaches may lead to different conclusions.
We illustrated the experimental uses of these measures with an existing RS and two well-known datasets.
We then presented a method to explore the effects of including different meta-paths in the recommendation process, evaluated its accuracy, and analyzed recommendations produced by combining different meta-paths.
We showed how including different parts of a HIN in a recommendation may significantly change accuracy and diversity.
For both datasets, including meta-paths with user-content information improved accuracy but often reduced collective diversity. 

Finally, we tested the influence of the semantic content on the accuracy and diversity of recommendations.
We randomized part of the HIN structure to remove its semantic meaning, while keeping the degree sequence of the HIN untouched.
Surprisingly, this random shuffling did not change significantly accuracy or diversity.
While these results are limited to a restricted number of datasets and parameters, they indicate that the effects of the information encoded in the structure of a HIN on recommendations are not only due to their semantic meaning.


\section*{Acknowledgements}

We warmly thank Elizabeth Doggett for her proofreading of the manuscript.
This work has been funded by the French National Agency of Research (ANR) under grants ANR-15-CE38-0001 (Algodiv), ANR-19-CE38-0006 (GOPI), and ANR LabCom FiT.

\bibliographystyle{plain}
\bibliography{references}

\begin{thebibliography}{10}

\bibitem{burke2002hybrid}
Robin Burke.
\newblock Hybrid recommender systems: Survey and experiments.
\newblock {\em User modeling and user-adapted interaction}, 12(4):331--370,
  2002.

\bibitem{burke2014hybrid}
Robin Burke, Fatemeh Vahedian, and Bamshad Mobasher.
\newblock Hybrid recommendation in heterogeneous networks.
\newblock In {\em Proceedings of UMAP'14}, pages 49--60. Springer, 2014.

\bibitem{dubey2011diversity}
Avinava Dubey, Soumen Chakrabarti, and Chiranjib Bhattacharyya.
\newblock Diversity in ranking via resistive graph centers.
\newblock In {\em Proceedings of SIGKDD'11}, pages 78--86. ACM, 2011.

\bibitem{harper2016movielens}
F~Maxwell Harper and Joseph~A Konstan.
\newblock The movielens datasets: History and context.
\newblock {\em TIIS}, 5(4):19, 2016.

\bibitem{hatami2014improving}
Majid Hatami and Saeid Pashazadeh.
\newblock Improving results and performance of collaborative filtering-based
  recommender systems using cuckoo optimization algorithm.
\newblock {\em IJCA}, 88(16), 2014.

\bibitem{herlocker2004evaluating}
Jonathan~L Herlocker, Joseph~A Konstan, Loren~G Terveen, and John~T Riedl.
\newblock Evaluating collaborative filtering recommender systems.
\newblock {\em TOIS}, 22(1):5--53, 2004.

\bibitem{hurley2011novelty}
Neil Hurley and Mi~Zhang.
\newblock Novelty and diversity in top-n recommendation--analysis and
  evaluation.
\newblock {\em TOIT}, 10(4):14, 2011.

\bibitem{jiang2019degenerate}
Ray Jiang, Silvia Chiappa, Tor Lattimore, Andras Agyorgy, and Pushmeet Kohli.
\newblock Degenerate feedback loops in recommender systems.
\newblock {\em arXiv:1902.10730}, 2019.

\bibitem{kunaver2017diversity}
Matev{\v{z}} Kunaver and Toma{\v{z}} Po{\v{z}}rl.
\newblock Diversity in recommender systems--a survey.
\newblock {\em Knowledge-Based Systems}, 123:154--162, 2017.

\bibitem{lhuillier2016new}
Amaury L'Huillier, Sylvain Castagnos, and Anne Boyer.
\newblock The new challenges when modeling context through diversity over time
  in recommender systems.
\newblock In {\em Proceedings of UMAP'16}, pages 341--344. ACM, 2016.

\bibitem{li2012scalable}
Rong-Hua Li and Jeffery~Xu Yu.
\newblock Scalable diversified ranking on large graphs.
\newblock {\em TKDE}, 25(9):2133--2146, 2012.

\bibitem{mcnee2006being}
Sean~M McNee, John Riedl, and Joseph~A Konstan.
\newblock Being accurate is not enough: how accuracy metrics have hurt
  recommender systems.
\newblock In {\em CHI'06}, pages 1097--1101. ACM, 2006.

\bibitem{nandanwar2018fusing}
Sharad Nandanwar, Aayush Moroney, and M~Narasimha Murty.
\newblock Fusing diversity in recommendations in heterogeneous information
  networks.
\newblock In {\em Proceedings of WSDM'18}, pages 414--422. ACM, 2018.

\bibitem{newman2018networks}
Mark Newman.
\newblock The configuration model.
\newblock In {\em Networks}, chapter~12. Oxford university press, 2018.

\bibitem{pariser2011filter}
Eli Pariser.
\newblock {\em The filter bubble: How the new personalized web is changing what
  we read and how we think}.
\newblock Penguin, 2011.

\bibitem{ramaciotti2020measuring}
Pedro Ramaciotti-Morales, Robin Lamarche-Perrin, Raphael Fournier-S'niehotta,
  Remy Poulain, Lionel Tabourier, and Fabien Tarissan.
\newblock Measuring diversity in heterogeneous information networks.
\newblock {\em arXiv:2001.01296}, 2020.

\bibitem{ricci2011introduction}
Francesco Ricci, Lior Rokach, and Bracha Shapira.
\newblock {\em Recommender systems handbook}.
\newblock Springer, 2011.

\bibitem{shi2018heterogeneous}
Chuan Shi, Binbin Hu, Wayne~Xin Zhao, and Philip~S Yu.
\newblock Heterogeneous information network embedding for recommendation.
\newblock {\em TKDE}, 31(2):357--370, 2018.

\bibitem{shi2016survey}
Chuan Shi, Yitong Li, Jiawei Zhang, Yizhou Sun, and Philip~S Yu.
\newblock A survey of heterogeneous information network analysis.
\newblock {\em TKDE}, 29(1):17--37, 2016.

\bibitem{shi2017heterogeneous}
Chuan Shi and Philip~S. Yu.
\newblock {\em Heterogeneous information network analysis and applications}.
\newblock Springer, 2017.

\bibitem{shi2015semantic}
Chuan Shi, Zhiqiang Zhang, Ping Luo, Philip~S Yu, Yading Yue, and Bin Wu.
\newblock Semantic path based personalized recommendation on weighted
  heterogeneous information networks.
\newblock In {\em Proceedings of CIKM'15}, pages 453--462. ACM, 2015.

\bibitem{silveira2019good}
Thiago Silveira, Min Zhang, Xiao Lin, Yiqun Liu, and Shaoping Ma.
\newblock How good your recommender system is? a survey on evaluations in
  recommendation.
\newblock {\em IJMLC}, 10(5):813--831, 2019.

\bibitem{stirling2007general}
Andy Stirling.
\newblock A general framework for analysing diversity in science, technology
  and society.
\newblock {\em J. R. Soc. Interface}, 4(15):707--719, 2007.

\bibitem{sun2011pathsim}
Yizhou Sun, Jiawei Han, Xifeng Yan, Philip~S Yu, and Tianyi Wu.
\newblock Pathsim: Meta path-based top-k similarity search in heterogeneous
  information networks.
\newblock {\em PVLDB}, 4(11):992--1003, 2011.

\bibitem{tang2012cross}
Jie Tang, Sen Wu, Jimeng Sun, and Hang Su.
\newblock Cross-domain collaboration recommendation.
\newblock In {\em Proceedings of SIGKDD'12}, pages 1285--1293. ACM, 2012.

\bibitem{tong2011diversified}
Hanghang Tong, Jingrui He, Zhen Wen, Ravi Konuru, and Ching-Yung Lin.
\newblock Diversified ranking on large graphs: an optimization viewpoint.
\newblock In {\em Proceedings of SIGKDD'11}, pages 1028--1036. ACM, 2011.

\bibitem{yang2012circle}
Xiwang Yang, Harald Steck, and Yong Liu.
\newblock Circle-based recommendation in online social networks.
\newblock In {\em Proceedings of SIGKDD'12}, pages 1267--1275. ACM, 2012.

\bibitem{yu2014personalized}
Xiao Yu, Xiang Ren, Yizhou Sun, Quanquan Gu, Bradley Sturt, Urvashi Khandelwal,
  Brandon Norick, and Jiawei Han.
\newblock Personalized entity recommendation: A heterogeneous information
  network approach.
\newblock In {\em Proceedings of WSDM'14}, pages 283--292. ACM, 2014.

\bibitem{yu2013recommendation}
Xiao Yu, Xiang Ren, Yizhou Sun, Bradley Sturt, Urvashi Khandelwal, Quanquan Gu,
  Brandon Norick, and Jiawei Han.
\newblock Recommendation in heterogeneous information networks with implicit
  user feedback.
\newblock In {\em Proceedings of RecSys'13}, pages 347--350. ACM, 2013.

\bibitem{Zafarani+Liu:2009}
R.~Zafarani and H.~Liu.
\newblock Social computing data repository at {ASU}, 2009.

\bibitem{zhang2002novelty}
Yi~Zhang, Jamie Callan, Jamie Callan, and Thomas Minka.
\newblock Novelty and redundancy detection in adaptive filtering.
\newblock In {\em Proceedings of SIGIR'02}, pages 81--88. ACM, 2002.

\bibitem{zhou2010solving}
Tao Zhou, Zolt{\'a}n Kuscsik, Jian-Guo Liu, Mat{\'u}{\v{s}} Medo,
  Joseph~Rushton Wakeling, and Yi-Cheng Zhang.
\newblock Solving the apparent diversity-accuracy dilemma of recommender
  systems.
\newblock {\em PNAS}, 107(10):4511--4515, 2010.

\bibitem{ziegler2005improving}
Cai-Nicolas Ziegler, Sean~M McNee, Joseph~A Konstan, and Georg Lausen.
\newblock Improving recommendation lists through topic diversification.
\newblock In {\em Proceedings of WWW'05}, pages 22--32. ACM, 2005.

\end{thebibliography}

\end{document}